\documentclass[floatfix,superscriptaddress,showpacs,amssymb,10pt,aps,prd,reprint,longbibliography]{revtex4-2}
\usepackage{placeins}
\usepackage{graphicx,epsfig,amssymb}
\usepackage{amsmath,amsfonts,times}
\usepackage{bm}

\usepackage[linktocpage,colorlinks]{hyperref}
\usepackage[caption=false]{subfig}
\usepackage[usenames]{color}
\usepackage{float}
\definecolor{coolblack}{rgb}{0.0, 0.18, 0.39}
\definecolor{darkred}{rgb}{0.5,0,0}
\definecolor{darkgreen}{rgb}{0,0.5,0}
\definecolor{darkblue}{rgb}{0,0,0.5}
\definecolor{lapislazuli}{rgb}{0.15, 0.38, 0.61}
\definecolor{venetianred}{rgb}{0.78, 0.03, 0.08}
\definecolor{bleudefrance}{rgb}{0.19, 0.55, 0.91}
\hypersetup{colorlinks=true, citecolor=darkgreen, linkcolor=darkblue, urlcolor=blue}
\graphicspath{{images/}}

\renewcommand{\vec}[1]{\boldsymbol{#1}}
\newcommand{\dd}{\mathrm{d}}
\newcommand{\ee}{\operatorname{e}}

\usepackage[nameinlink,noabbrev]{cleveref}
\crefname{equation}{Eq.}{Eqs.}
\Crefname{equation}{Equation}{Equations}
\crefname{figure}{Fig.}{Figs.}
\Crefname{figure}{Fig.}{Figs.}
\crefformat{figure}{Fig.~#2#1#3}
\Crefformat{figure}{Fig.~#2#1#3}
\begin{document}
\author{Junyi Li }
\address{School of Physics and Technology, Wuhan University, Wuhan, 430072, China}
\author{Yinrong Fang}
\address{School of Physics and Technology, Wuhan University, Wuhan, 430072, China}
\author{Yang Huang}
\address{School of Physics and Technology, University of Jinan,
336 West Road of Nan Xinzhuang, Jinan, Shandong 250022, China}
\author{Qian Li }
\email[Corresponding author:~]{liqiann@whu.edu.cn}
\address{School of Physics and Technology, Wuhan University, Wuhan, 430072, China}
\author{Junji Jia }
\email[Corresponding author:~]{junjijia@whu.edu.cn}
\address{Department of Astronomy $\&$ MOE Key Laboratory of Artificial Micro- and Nano-structures, School of Physics and Technology, Wuhan University, Wuhan, 430072, China}

\title{\Large Scattering of scalar waves by scalarized black holes in Einstein-scalar-Gauss-Bonnet gravity}

\begin{abstract}
We systematically investigate the scattering and absorption of a massive scalar field by scalarized black holes in Einstein-scalar-Gauss-Bonnet gravity with quartic coupling \(f(\phi)=\alpha(\phi^{2}+\beta\phi^{4})\) and exponential coupling \(f(\phi)=\alpha(1-\ee^{-\beta\phi^{2}})\). The background solutions are obtained numerically via a Chebyshev pseudospectral method, enabling a partial-wave analysis on purely numerical black hole backgrounds. The partial-wave results are cross-validated against geodesic and glory approximations and the Schwarzschild limit with excellent agreements. We find that scalarization suppresses the total absorption cross section relative to the Schwarzschild case, and produces distinctive signatures in the differential scattering cross section. A qualitative difference between the two coupling models is identified: for the quartic coupling model, deviations from Schwarzschild are concentrated in the intermediate-angle region \(\sim(70^{\circ},140^{\circ})\), with comparatively small changes at forward angles and near the glory peak; for the exponential coupling model, deviations appear across the full angular range and substantially modify the glory peak, which in some parameter regimes can even exceed the Schwarzschild value. This distinctive angular pattern provides a potential diagnostic for distinguishing EsGB coupling functions from scattering observables. 
\end{abstract}


\maketitle

\section{Introduction}\label{sec:intro}

The advent of gravitational-wave astronomy and black hole (BH) imaging has opened unprecedented opportunities to test general relativity (GR) in the strong-field regime~\cite{Yagi_2016}. To date, all systematic tests, from the LIGO-Virgo-KAGRA observing runs~\cite{PhysRevLett.116.221101,PhysRevD.100.104036,PhysRevD.112.084080} 
and the single-event analysis of GW250114~\cite{6c61-fm1n} to the Event Horizon Telescope (EHT) shadow observations of M87* and Sgr~A*~\cite{PhysRevLett.125.141104,pub.1147844694} have found no evidence for deviations from GR, implying that any beyond-GR effects must be subtle and their detection demands observables with enhanced discriminating power. The mainstream probes employed so far, including inspiral phase, quasinormal-mode (QNM) frequencies, and shadow observations, each encode partial information about the BH spacetime~\cite{Yagi_2016}. The differential scattering cross section \(d\sigma/d\Omega\) offers a complementary probe. It continuously encodes the radial structure of the effective potential \(V_{\mathrm{eff}}(r)\) across the full angular range, and the partial-wave interference pattern may translate small quantitative differences in the metric into qualitative changes in the fringe frequency and envelope shape, which are otherwise not readily distinguishable using a limited set of observables, such as individual QNM frequencies or shadow radii. 

Arising as a spontaneous scalarization of BHs, Einstein-scalar-Gauss-Bonnet (EsGB) gravity represents a valuable tool for black hole physics, since it features BHs with spontaneous scalar hair, and it is also a subclass of Horndeski gravity~\cite{Blázquez-Salcedo2024}. A central degree of freedom in the EsGB gravity is the choice of the coupling function \(f(\phi)\), which determines the scalar–Gauss–Bonnet interaction and parametrizes different possible realizations of the theory. In this work we consider two representative forms. The quartic coupling model (QCM) with \(f(\phi)=\alpha(\phi^{2}+\beta\phi^{4})\), representing a generic effective-field-theory truncation, and the exponential coupling model (ECM) with \(f(\phi)=\alpha(1-\ee^{-\beta\phi^{2}})\), motivated by the string-theoretic dilaton structure~\cite{PhysRevLett.120.131103,PhysRevLett.120.131104}. The fundamental scalarized solutions of both models have been shown to be mode-stable in certain regions of parameter space~\cite{PhysRevD.109.104057,PhysRevD.99.064011,PhysRevD.98.084011,PhysRevD.102.024086}, allowing the self-consistency of performing test-field scattering calculations on these backgrounds.

The strong-field properties of scalarized EsGB BHs have been extensively studied. Inspiral-merger-ringdown waveform models have been constructed~\cite{PhysRevD.111.024016}, nonperturbative QNMs have been calculated~\cite{sq4g-mmxp}, and EHT shadow data have been used to constrain the coupling parameter~\cite{Wang2024}. Yet the scattering cross section---despite its proven diagnostic power in many famous spacetimes, including Schwarzschild~\cite{Futterman1988}, Reissner-Nordstr\"om~\cite{PhysRevD.79.064022}, Kerr~\cite{KostasGlampedakis_2001}, Kerr-Newman~\cite{Li2026}
as well as in parametrized~\cite{Magalhaes2022}  
and charged Horndeski~\cite{PhysRevD.111.024059} BH spacetimes---has never been applied to scalarized EsGB BHs. One of the reasons for this is the technical challenge that the metrics are only numerically available for the EsGB BHs. In this work, we overcome this technical challenge and study the scattering of a massive scalar field in the EsGB background. We present the first systematic computation of scalar-field scattering cross sections for spontaneously scalarized EsGB BHs, covering both the quartic and exponential coupling models. Our calculations of the partial-wave scattering here are performed on a purely numerical black hole metric. We validate the computation algorithsm and results through three independent consistency checks, exact recovery of the Schwarzschild limit, agreement with the geodesic geometric-optics approximation~\cite{FORD1959259}, and agreement with the glory approximation~\cite{PhysRevLett.52.2313,PhysRevD.31.1869}.

Our central physical findings are twofold. First, the scattering cross sections of both models deviate significantly from the Schwarzschild result. The interference fringe frequency, envelope shape, and the height of glory peak are all visibly modified, establishing the scattering cross section as a viable test of whether EsGB modifications are present. Second, relative to the Schwarzschild baseline, the two coupling functions produce qualitatively distinct angular-distribution patterns. the ECM exhibits deviations from the Schwarzschild result across the full angular range including around the glory peak, whereas the QCM shows deviations from the Schwarzschild result concentrated in the intermediate-angle region \(\sim(70^{\circ},140^{\circ})\) with comparatively little change at small angles and near the glory. This angular fingerprint may provide a useful diagnostic among coupling function types within EsGB.

The remainder of this paper is organized as follows. Section~\ref{sec:background} introduces the EsGB action and the two coupling function models, and presents the corresponding numerical BHs. Section~\ref{sec:scattering} presents the scattering formalism, including the partial-wave method and the classical effective potential. Subsection~\ref{sec:geo} presents the geodesic analysis, including geometric absorption, classical scattering, and glory scattering. Section~\ref{sec:results} presents the numerical absorption and differential scattering cross sections and compares the partial-wave results with the geometric and glory approximations. In the entire work, the background metrics we used are chosen from parameter ranges in which they are stable against the relevant perturbations and we neglect the backreaction of the scalar field on the background geometry.  We summarize the main findings and discuss future directions in Section \ref{sec:conclusion}. Throughout this paper we use natural units \(G=c=\hbar=1\) and metric signature \((-,+,+,+)\).

\section{EsGB gravity and solution method}\label{sec:background}

\subsection{EsGB gravity models}

We start with the following action describing a general class of higher-curvature gravitational theories
\begin{equation}
	S=\frac{1}{16\pi}\int \dd^4x\sqrt{-g}\left[R - \frac{1}{2}(\partial \phi)^2 + f(\phi)\mathcal{G}\right],\label{eq:action}
\end{equation}
where the matter field \(\phi\) is coupled to the Gauss-Bonnet term \(\mathcal{G}\) through the coupling function \(f(\phi)\). Here, \(g\equiv \det(g_{\mu\nu})\), and \(R\) is the Ricci scalar. \((\partial\phi)^2\equiv g^{\mu\nu}\partial_\mu\phi\partial_\nu\phi\) with \(\alpha\) being the coupling constant. 
The Gauss-Bonnet term is defined as
\begin{equation}
\mathcal{G}=R_{\mu\nu\rho\sigma}R^{\mu\nu\rho\sigma}-4R_{\mu\nu}R^{\mu\nu}+R^2.\label{eq:GB}
\end{equation}
Variation of this action yields the Einstein's equation
\begin{equation}
	G_{\mu\nu}=T_{\mu\nu},\label{eq:einstein}
\end{equation}
and Klein-Gordon's equation
\begin{equation}
	\Box\phi+\dot f(\phi)\mathcal{G}=0.\label{eq:KG}
\end{equation}
Here the energy-momentum tensor \(T_{\mu\nu}\) is ~\cite{PhysRevD.97.084037}
\begin{align}
		T_{\mu\nu}
		=& -\frac{1}{4} g_{\mu\nu}\,\partial_\rho \phi\,\partial^\rho \phi
		+ \frac{1}{2}\,\partial_\mu \phi\,\partial_\nu \phi \nonumber\\
		&- \frac{1}{2}\left(g_{\rho\mu} g_{\lambda\nu}
		+ g_{\lambda\mu} g_{\rho\nu}\right)
		\eta^{\kappa\lambda\alpha\beta}
		\tilde{R}^{\rho\gamma}{}_{\alpha\beta}
		\nabla_\gamma \partial_\kappa f(\phi),\label{eq:stress-energy}
\end{align}
with the Hodge dual of the Riemann tensor as
\begin{equation}
	\tilde{R}^{\rho\gamma}{}_{\alpha\beta}
	= \eta^{\rho\gamma\sigma\tau} R_{\sigma\tau\alpha\beta}
	= \frac{\epsilon^{\rho\gamma\sigma\tau}}{\sqrt{-g}}
	\,R_{\sigma\tau\alpha\beta},\label{eq:dualR}
\end{equation}
where $\eta^{\rho\gamma\sigma\tau}$ is the Levi-Civita tensor while  $\epsilon^{\rho\gamma\sigma\tau}$ is the Levi-Civita symbol. In Eq. \eqref{eq:KG}  and hereafter, a dot represents the derivative with respect to \(\phi\). In this work, we will focus on static and spherically symmetric BH solutions and therefore adopt the following metric
\begin{equation}
	\dd s^2=-\ee^{A(r)}\dd t^2+\ee^{B(r)}\dd r^2+r^2\dd\theta^2+r^2\sin^2\theta\, \dd\varphi^2.\label{eq:metric}
\end{equation}
Substituting this metric and the scalar field \(\phi=\phi(r)\) into~\cref{eq:einstein,eq:KG}, we obtain a system of coupled ordinary differential equations (ODEs) for the metric functions \(A(r),B(r)\) and the scalar field \(\phi(r)\)
\begin{subequations}
\label{eq:odes}
\begin{align}
    &-4 B' \Bigl[ r \ee^B + 2(\ee^B - 3)  \dot f \phi' \Bigr]- 4(\ee^B - 1) \Bigl( \ee^B - 4  \dot f \phi'' \Bigr) \nonumber 
    \\
      &+(\phi')^2 \Bigl[ r^2 \ee^B
+ 16(\ee^B - 1)  \ddot f \Bigr]   = 0,\label{eq:ode1}\\
   & 1 - \ee^B + r A' - \frac{1}{4}\phi' \Bigl[ 8(3 \ee^{-B} - 1)  A' \dot f + r^2 \phi' \Bigr] = 0, \label{eq:ode2}\\
     &2  \dot f \Bigl[ -2(\ee^B - 1)(A')^2 + 2(\ee^B - 3)A' B'
     \nonumber 
     - 4(\ee^B - 1)A'' \Bigr]\\
     +& r \ee^B \Bigl[ (4 + r A' - r B')\phi' + 2r\phi'' \Bigr] = 0. \label{eq:ode3}
\end{align}
\end{subequations}
Here and henceforth, primes denote derivative with respect to \(r\).

\subsection{Boundary conditions and numerical techniques} 

We use the Chebyshev pseudospectral method to solve the coupled ODEs \eqref{eq:odes}. Let \(r_h\) denote the event horizon radius. We map the semi-infinite radial domain \([r_h,+\infty)\) onto the compact domain \([-1,+1]\) through the transformation
\begin{align}
	x=1-\frac{2r_h}{r}.
\end{align}
To handle the singular behavior of the metric function at the horizon, we also define two auxiliary functions $H(x)$ and $G(x)$ corresponding to $A(x)$ and $B(x)$ respectively, through
\begin{align}
	\ee^A\equiv H(x)\frac{1+x}{2},\quad \ee^B\equiv G(x)\frac{2}{1+x}.
\end{align}
From this point onward, all discussion in this subsection will be based on the \(x\)-coordinate and functions \(H(x)\) and \(G(x)\). 

For the boundary conditions, at infinity \(x=1\), the asymptotic flatness conditions impose
\begin{align}
	H(1)=1,\quad G(1)=1,\quad \phi(1)=0.\label{eq:bcl}
\end{align}

Requiring regularity of the scalar-field equation at the horizon \(x=-1\) yields the following condition on the horizon derivative
\cite{PhysRevLett.120.131103,PhysRevD.97.084037}
\begin{align}
	\partial_x\phi|_{x=-1} = \frac{r_h^2}{8 \dot{f_h}}\left(-1 \pm \sqrt{1 - \frac{96 (\dot{f}_h)^2}{r_h^{4}}}\right),\label{eq:phih}
\end{align}
where \(\dot f_h \equiv \dot f(\phi_h)\), with \(\phi_h\equiv \phi(x=-1)\). The existence of a real solution requires the discriminant 
\begin{align}
	\Delta = 1 - \frac{96 (\dot{f}_h)^2}{r_h^{4}}\label{discriminant}
\end{align}
to be nonnegative. In actual calculations, only the \(+\) branch of \(\partial_x\phi|_{x=-1}\) is physical since only in this case can we recover the trivial solution in the limit \(\phi_h\to 0\). On the other hand, the regularity of \(H\) on the horizon leads to the equation
\begin{align}
	\left. \big( \mathcal{C}_1 \partial_x H + \mathcal{C}_0 H \big) \right|_{x=-1} &= 0 ,\label{eq:h}
\end{align}
where the boundary coefficients are
\begin{align}
	\mathcal{C}_1 =&~ 2 G \big( r_h^2 + 4 \dot{f} \partial_x \phi \big) , \notag \\
	\mathcal{C}_0 =& - G \big[ G + 2 \partial_x G + (\partial_x \phi)^2 - 1 \big] r_h^2 \notag \\
	&- 4 \big[ 3 \dot{f} \partial_x \phi (1 + G) - 2 G \big( \ddot{f} (\partial_x \phi)^2 + \dot{f} \partial_x^2 \phi \big) \big] .
\end{align}
Here, the boundary value of the spatial metric function $G(x)$ at the horizon $x=-1$ is not a free parameter. It is algebraically fixed by eliminating the leading-order pole of~\cref{eq:ode2} at the horizon, which yields
\begin{align}
	G(-1) = 1 + \frac{4  \dot{f}_h}{r_h^2} \left. \phi_x \right|_{x=-1} . \label{eq:g}
\end{align}

After imposing the boundary conditions described above, we discretize the unknown functions \(H(x)\), \(G(x)\), and \(\phi(x)\) on the Chebyshev-Gauss-Lobatto collocation points
\begin{align}
    x_j=\cos\left(\frac{j\pi}{N}\right),\qquad j=0,1,\ldots,N .
\end{align}
In the calculations presented below we will use $N=19$.
The numerical unknowns are the nodal values
\begin{align}
    \{H_j,G_j,\phi_j\}_{j=0}^{N},
\end{align}
where $ H_j=H(x_j),\quad G_j=G(x_j),\quad \phi_j=\phi(x_j)$. Derivatives with respect to \(x\) are evaluated by the Chebyshev differentiation matrix~\cite{boyd2013chebyshev}. 

Substituting the nodal values and their spectral derivatives into the field equations and boundary conditions gives a nonlinear algebraic system with \(3(N+1)\) unknowns. Denoting the full vector of unknowns by
\begin{align}
    \mathbf{U}    =    (H_0,\ldots,H_N,G_0,\ldots,G_N,\phi_0,\ldots,\phi_N)^{T},
\end{align}
the discretized system can be written as
\begin{align}
    \mathbf{R}(\mathbf{U})=\mathbf{0}.
\end{align}
Here $\mathbf{R}(\mathbf{U})$ denotes the vector function constructed from the $3(N-1)$ ODEs~\cref{eq:odes} and 6 boundary conditions~\cref{eq:bcl,eq:phih,eq:h,eq:g}, covering the full \(3(N+1)\) unknowns. This system is solved by a damped Newton iteration method
\begin{align}
    \mathbf{J}^{(k)}\delta\mathbf{U}^{(k)}
    =
    -\mathbf{R}(\mathbf{U}^{(k)}),
\,
    \mathbf{U}^{(k+1)}
    =
    \mathbf{U}^{(k)}
    +
    \lambda^{(k)}\delta\mathbf{U}^{(k)}, 
    \label{eq:des}
\end{align}
where \(0<\lambda^{(k)}\leq 1\) is a damping factor chosen by backtracking line search to reduce the residual, $k$ denotes the iteration counter and the Jacobian matrix $\mathbf{J}$ is evaluated by a central finite-difference approximation~\cite{boyd2013chebyshev}. 

To start the calculation, we give the initial guess chosen as
\begin{align}
    H_j=1,\qquad
    G_j=1,\qquad
    \phi_j=\frac{1-x_j}{2}\phi_h^{\mathrm{init}},
\end{align}
and then the damped Newton iteration method yields the nodal values of the background functions. 
Since the numerical solution of the wave equations in the following section, such as \cref{eq:radial}, requires high-precision evaluation of the metric functions at arbitrary radii, we still need an accurate reconstruction scheme for the metric and field functions that preserves their spectral accuracy across the entire radial domain. To reconstruct continuous metric functions from the nodal values, we employ the barycentric interpolation formula~\cite{doi:10.1137/S0036144502417715}. For a general function \(u(x)\) with nodal values \(u_j=u(x_j)\), the barycentric interpolant is written as
\begin{align}
	p_N(x)=
	\frac{\displaystyle\sum_{j=0}^N \frac{w_j}{x-x_j} u_j}
	{\displaystyle\sum_{j=0}^N \frac{w_j}{x-x_j}},\label{eq:bary}
\end{align}
where \(x_j\) are the Chebyshev--Gauss--Lobatto nodes, \(w_j\) are the corresponding barycentric weights. The metric derivatives are obtained by first applying the Chebyshev differentiation matrix to the nodal values in spectral space, then interpolating the resulting derivative nodal values via \cref{eq:bary}, thereby preserving spectral accuracy in both the metric functions and their derivatives. This method avoided the local errors and derivative discontinuities introduced by conventional finite-difference and spline interpolation method, which might disrupt the phase extraction of the wave equation in the far field in later calculations.

\subsection{Models and their BH solutions}
\label{sec:metric}

In this section, we first specify the models and then present their corresponding solutions to the metric functions and scalar fields obtained using the boundary conditions~\cref{eq:bcl,eq:phih,eq:h,eq:g} and numerical techniques discussed in the previous subsection. We will compare these solutions with those of the Schwarzschild solution and the shape of the scalar field between the two models.

\subsubsection{QCM}
Spontaneous scalarization requires the coupling function to satisfy $\dot{f}(0)=0$ and $\ddot{f}(0)\neq 0$~\cite{PhysRevLett.120.131103,PhysRevLett.120.131104}. The simplest form satisfying these two conditions is the quadratic coupling $f\propto\phi^2$, but scalarized black hole solutions with a purely quadratic coupling are unstable under radial perturbations~\cite{PhysRevD.99.064011}. The inclusion of the quartic term $\beta\phi^4$ can stabilize the scalarized solutions~\cite{PhysRevD.99.044017}, where $\beta$ controls the strength of the higher-order suppression that quenches the tachyonic instability triggered by the quadratic term. Motivated by these considerations, we adopt the quartic coupling function
\begin{align}
	f(\phi)=\alpha(\phi^2+\beta \phi^4),
	\label{Poly-Modle}
\end{align}
which provides a simple construction satisfying the conditions for spontaneous scalarization. Here, $\alpha$ controls the overall coupling strength between the scalar field and the Gauss-Bonnet invariant. The stability of the BH solution of this model has been carefully studied in~\cite{PhysRevD.109.104057,PhysRevD.99.064011}, leading to the conclusion that the solution is stable in the following range of parameters 
\begin{align}
	\quad \beta\le-1,\quad -\frac{1}{2}<\beta \phi_h^2<-0.1155.
\end{align}
In addition, the discriminant~\cref{discriminant} for this model requires that
\begin{align}
	1-\frac{384\alpha^2\phi_h^2+1536\alpha^2\beta\phi_h^4+1536\alpha^2\beta^2\phi_h^6}{r_h^4}
	\ge 0.
	\label{condition_phi_Poly}
\end{align}

\begin{figure*}[htp!]
 \centering
\includegraphics[width=0.32\textwidth]{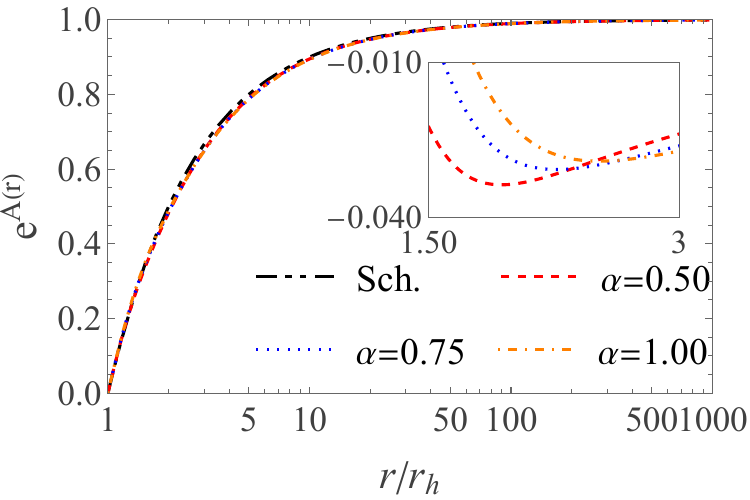}
\hfill
\includegraphics[width=0.33\textwidth]{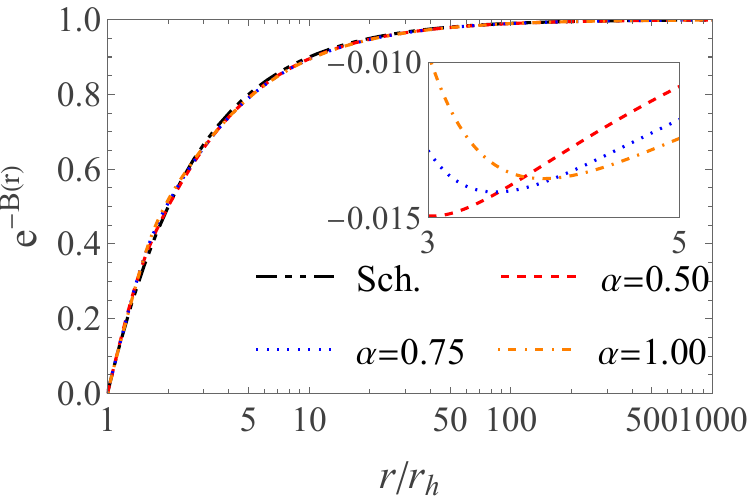}
\hfill
\includegraphics[width=0.33\textwidth]{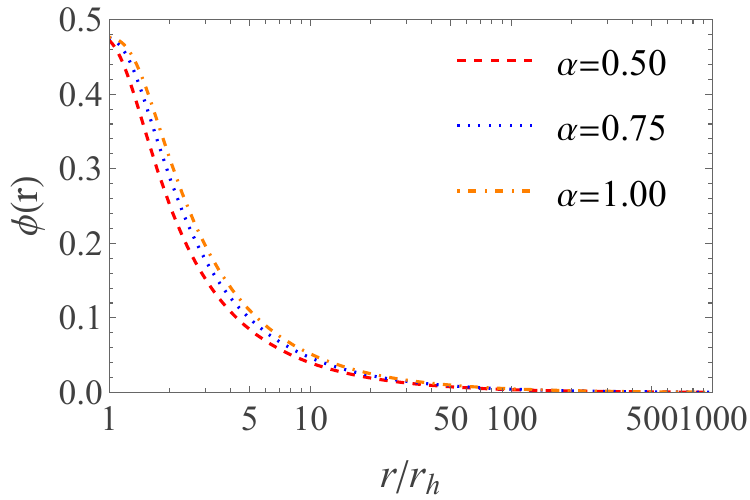}
\hfill
\includegraphics[width=0.33\textwidth]{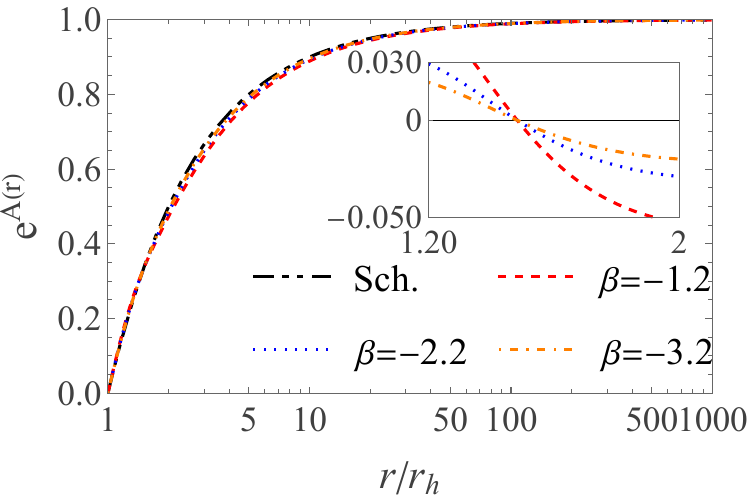}
\hfill
\includegraphics[width=0.33\textwidth]{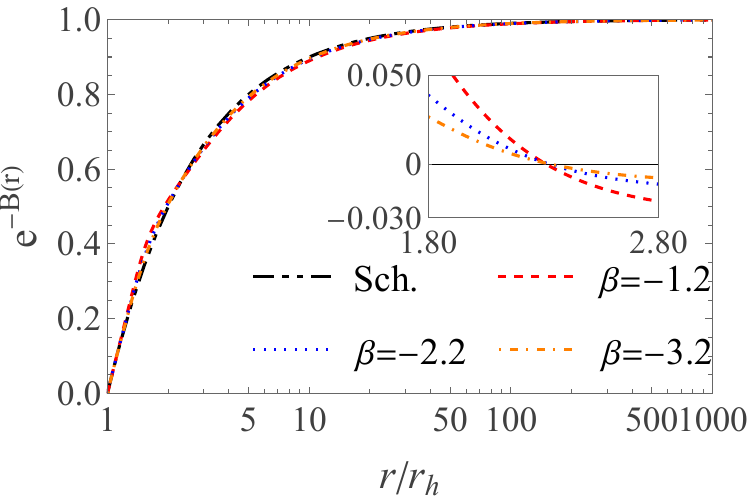}
\hfill
\includegraphics[width=0.33\textwidth]{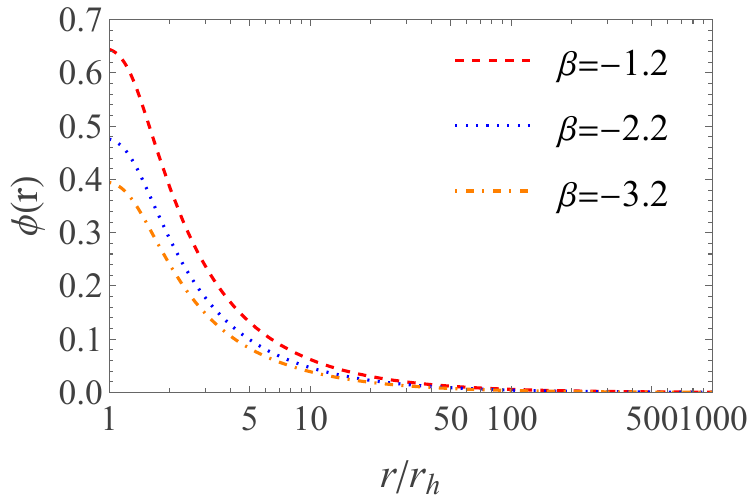}

\caption{Metric functions $-g_{tt}\equiv \ee^{A}$  (left column), $1/g_{rr}\equiv \ee^{-B}$ (middle column) and scalar field (right column) of the QCM. Top row: $\alpha=0.5,\,0.75,\,1.0$ and $\beta=-2.2$; Bottom row: $\alpha=0.75$ and $\beta=-1.2,\,-2.2,\,-3.2$.\label{fig:modelA}}
\end{figure*}
We then choose typical parameter combinations $(\alpha,\beta)$ respecting the above conditions and solve for the corresponding metric and field functions. These solutions are shown in Fig. \ref{fig:modelA}.
From the first row of Fig. \ref{fig:modelA}, it is seen that in the given parameter range, the difference between the QCM and the Schwarzschild BH becomes more significant as \(\alpha\) increases, but the scalar field profile does not change significantly.
To further quantify the impact of the coupling parameters on the global spacetime geometry, we examine the ADM mass of the three solutions displayed in Fig.\ref{fig:modelA}. For the fixed value $\beta=-2.2$, the masses corresponding to $\alpha=0.5,\,0.75,$ and $1$ are $M\approx 1.052M_s,\, 1.060M_s,\,$ and $1.066M_s$ respectively, where $M_s=0.5$ represents the mass of a Schwarzschild BH with the same $r_h=1$. This indicates that, much like the local metric deformations, the global mass of the black hole is only weakly sensitive to $\alpha$. 

On the other hand, as shown in the second row of Fig. \ref{fig:modelA}, the scalar field profile changes more dramatically as \(|\beta|\) increases, while the deviation of the metric functions from the Schwarzschild ones decreases. These effects of the parameters $\alpha$ and $\beta$ are consistent with the discussion in~\cite{PhysRevD.99.064011}, namely that quadratic terms can trigger instability while higher-order terms suppress the instability, which is the role of \(\beta\). The effect is also obvious in the ADM mass of each solution. For the fixed value $\alpha=0.75$, the masses corresponding to $\beta=-1.2,\,-2.2,$ and $-3.2$ are $M\approx 1.111M_s,\, 1.060M_s,\,$ and $1.042M_s$. These values show that the ADM mass decreases monotonically with increasing $|\beta|$, closely following the trend of the metric deviation. 

\subsubsection{ECM}

The exponential coupling function
\begin{align}\label{Exp-Modle}
	f(\phi)=\alpha(1-\ee^{-\beta\phi^2})
\end{align}
is inspired by the exponential dilaton--Gauss-Bonnet coupling $f\sim \ee^{\gamma\phi}$ arising in the low-energy effective action of string theory~\cite{PhysRevD.54.5049}, with $\phi$ replaced by $\phi^2$ in the exponent to impose the $\phi\to-\phi$ symmetry required for spontaneous scalarization~\cite{PhysRevLett.120.131103}. The constant term does not affect the field equations, since a constant multiplied by $\mathcal{G}$ is a topological invariant in four dimensions.
For this coupling function, stable scalarized solutions exist whenever the scalar field decreases monotonically from the horizon to spatial infinity. We only need to ensure the monotonic decrease of the scalar field \(\phi\) and the satisfaction of the discriminant condition, i.e.~\cref{discriminant},
\begin{align}
	1-\frac{384\alpha^2\beta^2\phi_h^2 \ee^{-2\beta\phi_h^2}}{r_h^4}
	\ge 0 .\label{eq:condition_phi_Exp}
\end{align}

\begin{figure*}[htp!]
 \centering
\includegraphics[width=0.32\textwidth]{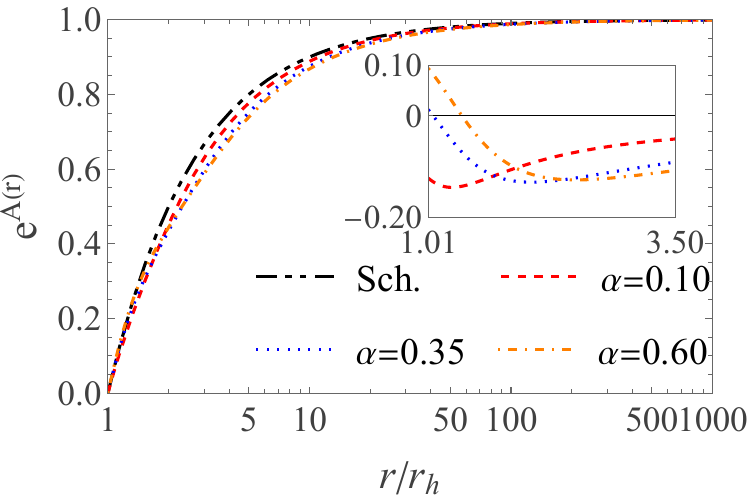}
\hfill
\includegraphics[width=0.33\textwidth]{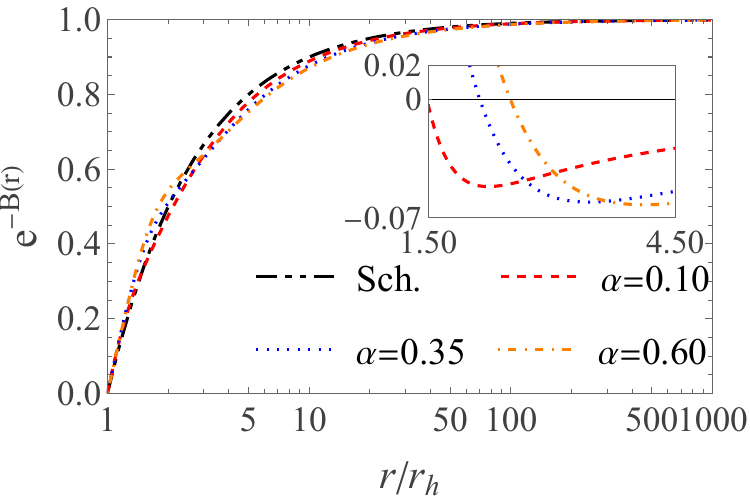}
\hfill
\includegraphics[width=0.33\textwidth]{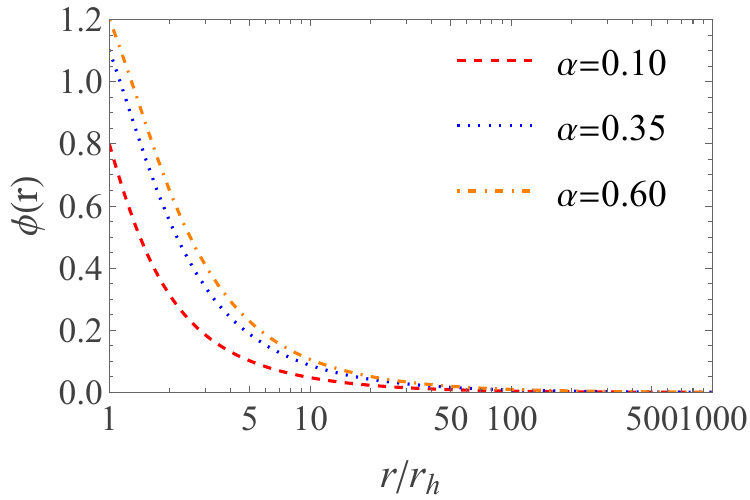}
\hfill
\includegraphics[width=0.33\textwidth]{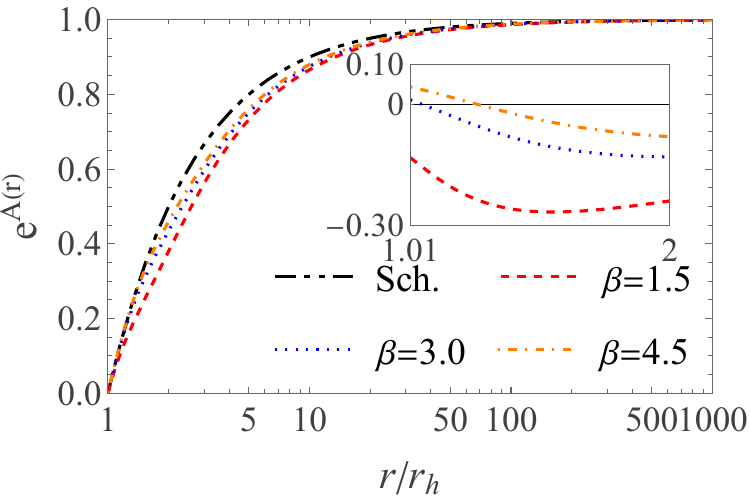}
\hfill
\includegraphics[width=0.33\textwidth]{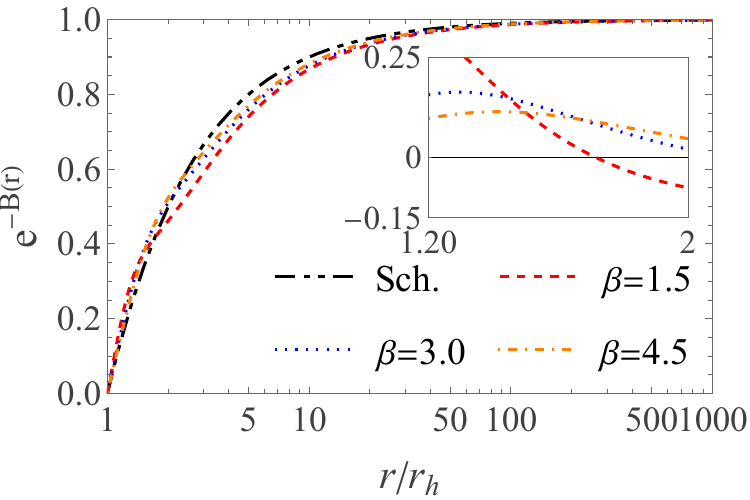}
\hfill
\includegraphics[width=0.33\textwidth]{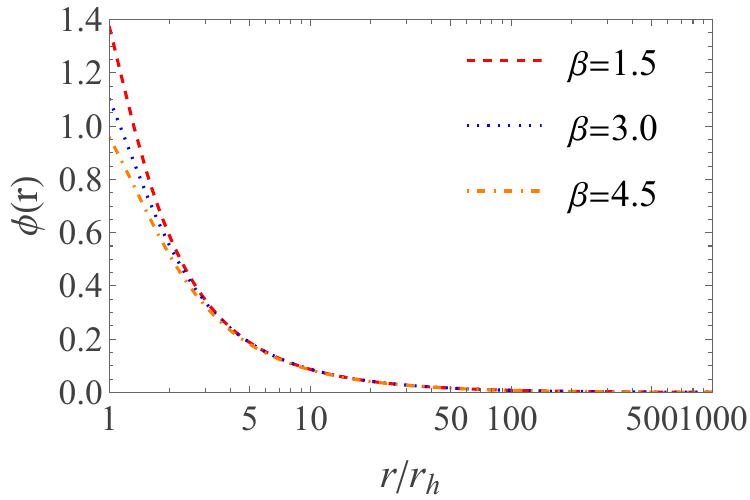}

\caption{Metric functions $-g_{tt}$ (left column), $1/g_{rr}$ (middle column) and scalar field (right column) of the ECM. Top row: $\alpha=0.1,\,0.35,\,0.6$ and $\beta=3.0$; Bottom row: $\alpha=0.35$ and $\beta=1.5,\,3.0,\,4.5$.\label{fig:modelB}}
\end{figure*}

Similar to the previous subsection, we also choose representative parameters $(\alpha,\,\beta)$ satisfying the above condition and then use the method described in Sec. \ref{sec:background} to solve for the corresponding metric and field functions. These solutions are presented in Fig. \ref{fig:modelB}. 

As shown in the first row, the difference between the ECM and the Schwarzschild BH metric functions becomes more apparent as \(\alpha\) increases, just as in the QCM seen from the top row of Fig. \ref{fig:modelA}. In addition, for the parameter values considered, the scalar-field profile is more sensitive to changes in \(\alpha\) than in the corresponding QCM examples. Here, for the fixed value $\beta=3.0$, the masses corresponding to $\alpha=0.1,\,0.35,$ and $0.6$ are $M\approx 1.116 M_s,\, 1.250 M_s,\,$ and $1.319M_s$.

For the effect of $\beta$, it is seen from the bottom plots that for the given parameter range, the deviation of the metric functions from the Schwarzschild case is slightly stronger than that as $\alpha$ changes. In contrast, the horizon value of the scalar field decreases as significantly as that in the QCM case when \(\beta\) increases. In this case, for fixed $\alpha=0.35$, the ADM masses of $\beta=1.5,\,3.0,\,4.5$ are $M\approx 1.342M_s,\, 1.250M_s,\,$ and $1.202M_s$. Meanwhile, the scalar field in the right column is not so sensitive to $\beta$, with $\phi_h$ decreasing monotonically as $\beta$ increases but the three curves soon converging at $r/r_h\gtrsim 2$. That means in ECM, changes in $\alpha$ have a greater impact on the scalar field than the changes in $\beta$.

\section{Absorption and scattering of a scalar field}\label{sec:scattering}
In this section, we study, using both the partial wave and geodesic methods, the absorption and scattering of a minimally coupled massive test scalar field $\Phi_\omega$ with a known frequency $\omega$ by the numerical BH spacetimes obtained for the QCM and ECM in the last section. We emphasize that it is of pivotal importance that the test scalar field $\Phi_{\omega}$ will not destabilize the metric because we have chosen the parameters $(\alpha,\,\beta)$ such that the BH solution is stable against scalar perturbations.

\subsection{Partial wave method} \label{sec:partial}
For this massive scalar field, its propagation satisfies the Klein-Gordon equation
\begin{align}\label{eq:KG_1}
	(\Box-\mu^2)\Phi_\omega=0,
\end{align}
where $\mu$ is the mass of the scalar field. Since the numerical BH is static and spherically symmetric, we consider a wave incoming from the $z^+$ axis so that we can directly use the following separable ansatz
\begin{align}\label{eq:ansatz}
	\Phi_{\omega } =\sum_{l=0}^{\infty}\frac{\psi_{\omega l}(r)}{r} P_l(\cos\theta) \ee^{-i \omega t},
\end{align}
where $P_l(\cos\theta)$ is the Legendre polynomial. Substituting it into Eq. \eqref{eq:KG_1}, we obtain the radial equation
\begin{align}\label{eq:radial}
	\ee^{A-B}\,\psi''_{\omega l}
	+\frac12\,\ee^{A-B}\,(A'-B')\,\psi'_{\omega l}
	+\left(\omega^2- V_{\text{eff}} \right)\psi_{\omega l}=0,
\end{align}
where $V_{\text{eff}}$ is the effective potential
\begin{align}\label{eq:veff}
	V_{\text{eff}}= e^{A} \left[
	\frac{l(l+1)}{r^2} + \mu^2 + \ee^{-B} \frac{A' - B'}{2r}
	\right].
\end{align}
Defining a tortoise coordinate $r_*$, i.e.,
\begin{align}
	\frac{\mathrm{d}r_*}{\mathrm{d}r}=\ee^{(B-A)/2},
\end{align}
Eq. \eqref{eq:radial} takes the Schr\"odinger-like form
\begin{align}\label{eq:radial_1}
	\frac{\dd^2\psi_{\omega l}}{\dd r_*^2}	+\left(\omega^2- V_{\text{eff}} \right)\psi_{\omega l}=0.
\end{align}

For scalar-field scattering by a BH, the following conditions on the horizon and at spatial infinity should be imposed
\begin{align}
	\psi_{\omega l}\approx
	\begin{cases}
		\ee^{-i k_\infty r_*} + R_{\omega l}\, \ee^{i k_\infty r_*}
		& r_* \to +\infty \ (r \to \infty), \\
		T_{\omega l}\, \ee^{-i \omega r_*}
		& r_* \to -\infty \ (r \to r_h),
	\end{cases}\label{BC_coditions_1}
\end{align}
where $k_\infty=\sqrt{\omega^2-\mu^2}$. To ensure that the scattered waves can propagate to infinity, the condition $\omega>\mu$ must be satisfied.
In addition, $T_{\omega l}$ and $R_{\omega l}$ denote the transmission and reflection coefficients, respectively. For scalar-field scattering by BHs, these coefficients satisfy the energy conservation relation~\cite{PhysRevD.111.024059}
\begin{align}
	|R_{\omega l}|^2 + \frac{\omega}{k_\infty}|T_{\omega l}|^2 = 1.
\end{align}

For a plane wave incident on a spherically symmetric BH, the total absorption cross section is
\cite{Sanchez:1977si,Benone:2014qaa}
\begin{align}\label{eq:abs}
	\sigma_{\text{abs}}=\sum_{l}\sigma_l=\sum_{l}\frac{\pi}{k_\infty^2}(2l+1)(1-|R_{\omega l}|^2),
\end{align}
and the differential scattering cross section is \cite{Sanchez:1977vz}
\begin{align}\label{eq:dscs}
	\frac{\dd\sigma}{\dd\Omega}=|f(\theta)|^2,
\end{align}
where \(f(\theta)\) is the scattering amplitude
\begin{align}\label{eq:sa}
	f(\theta)= \frac{1}{2 i k_{\infty}}\sum_{l=0}^{\infty}(2l+1)\big[(-1)^{\,l+1} R_{\omega l} - 1\big] P_l(\cos\theta).
\end{align}

In the next section, we use Eqs.~\eqref{eq:abs} and \eqref{eq:dscs} to compute the absorption and scattering cross sections for the numerical EsGB BHs. In doing so, since in our previous solutions to the BH metrics, the ADM mass $M$ varies while the horizon radius was fixed among different BH solutions, we will introduce $M$-normalized dimensionless quantities to enable a meaningful comparison of cross sections. Specifically, spacetime quantities are rendered dimensionless by normalizing them with $1/M$, so that ratios such as $r/M$ become dimensionless, whereas the field parameters are scaled by $M$. Consequently, the absorption cross section $\sigma_{\text{abs}}/M^2$ and the differential scattering cross section $d\sigma/(M^2 d\Omega)$ are dimensionless physical quantities. 
The resulting dimensionless quantities allow comparisons between solutions with different background parameters $(\alpha,\,\beta)$ but the same BH mass $M$. In the Schwarzschild limit, these dimensionless quantities exhibit universal behavior, and therefore any deviation from this universal curve reflects genuine modifications of the near-horizon geometry.

\begin{figure*}[htp!]
\centering
\includegraphics[width=0.47\textwidth]{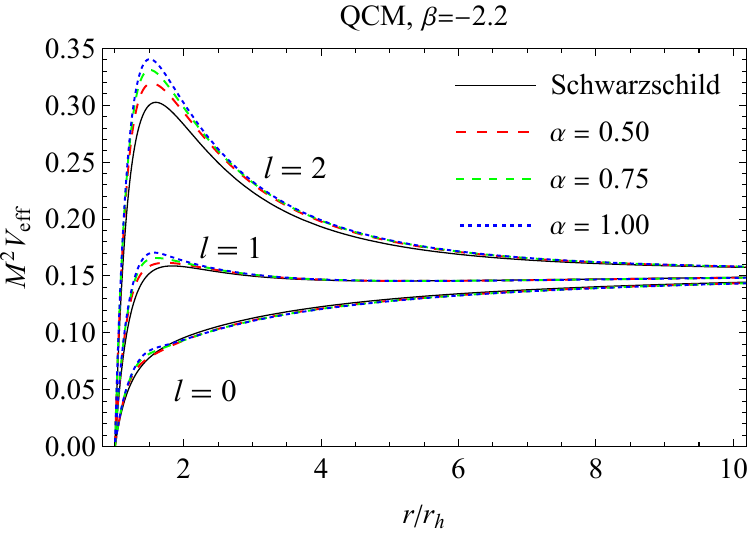}
\hfill
\includegraphics[width=0.47\textwidth]{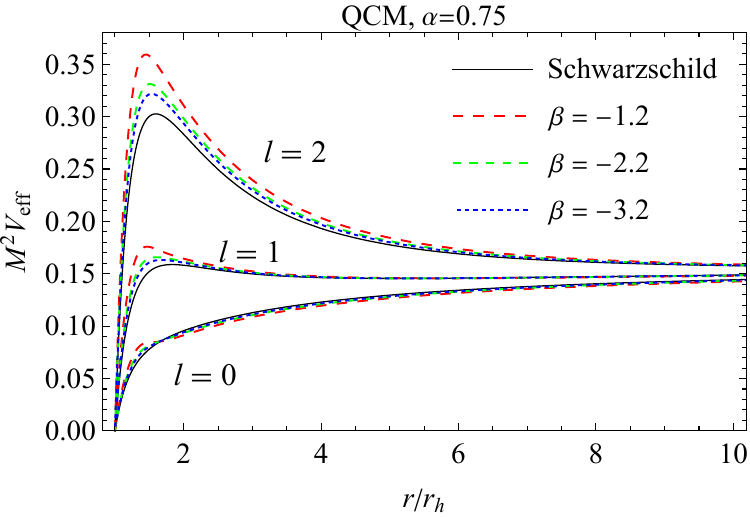}
\\
\includegraphics[width=0.47\textwidth]{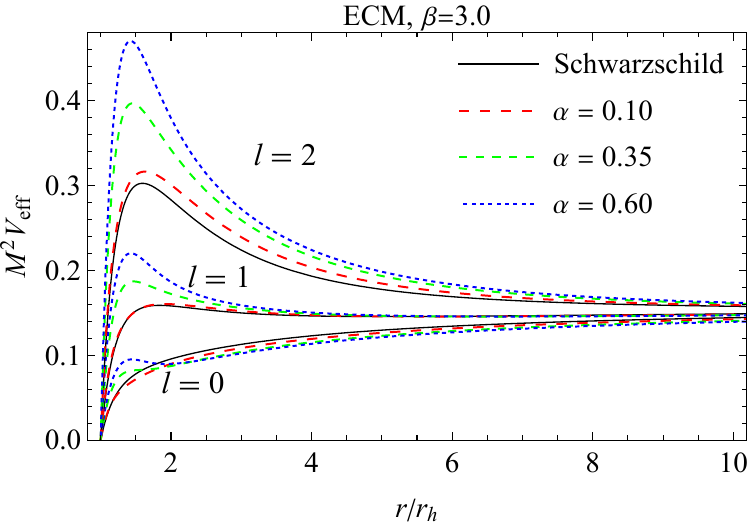}
\hfill
\includegraphics[width=0.47\textwidth]{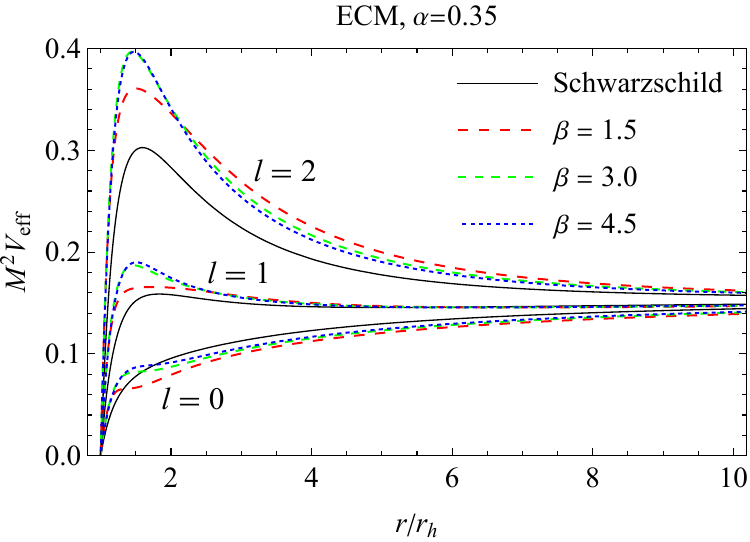}
	\caption{The effective potential $V_{\text{eff}}$ for the QCM (top row) and ECM (bottom row) with different coupling parameters. Top left (a): $\alpha=0.5,\,0.75,\,1.00$ and $\beta=-2.2$; Top right (b): $\alpha=0.75$ and $\beta=-1.2,\,-2.2,\,-3.2$. Bottom left (c): $\alpha=0.1,\,0.35,\,0.6$ and $\beta=3.0$; Bottom right (d): $\alpha=0.35$ and $\beta=1.5,\,3.0,\,4.5$. The other parameter is fixed to $M\mu=0.4$.}\label{fig:veff}
\end{figure*}

To better interpret the behavior of the absorption and scattering cross sections that will be shown later, here we plot in Fig.~\ref{fig:veff} the effective potential of the EsGB BH for different coupling parameters $\alpha$ and $\beta$ in both the QCM and ECM, where the effective potential~\cref{eq:veff}  has been rescaled as $M^2V_{\text{eff}}$ using the ADM mass of each BH, so that for any given $M\mu$, which is also dimensionless, the dimensionless effective potential has the same value $M^2\mu^2$ at infinity, which is helpful for our comparison. From Fig.~\ref{fig:veff}(a), we observe for partial waves with larger angular number ($l\geq2$) in the QCM, the peak of the effective potential with fixed $\beta$ increases as the coupling parameter $\alpha$ increases, indicating that the potential gradually deviates from that of the Schwarzschild BH. However, for fixed $\alpha$, as $|\beta|$ increases, the peak of the effective potential for $l\geq2$ partial waves decreases and gradually approaches that of the Schwarzschild BH, as shown in Fig.~\ref{fig:veff}(b). For $l=0$ and $l=1$ partial waves, however, the potential barrier does not necessarily vary monotonically with the coupling parameters $\alpha$ and $\beta$. From the expression for the effective potential in Eq. \eqref{eq:veff}, it is clear that this is a result of subtle competition between various terms involving the metric functions under these parameter settings. Anyway, in general, we see that as the angular momentum $l$ increases, the effect of both parameters $\alpha$ and $\beta$ becomes more pronounced. 

As for ECM, it can be seen from Fig.~\ref{fig:veff}(c) that for fixed $\beta$, the potential barrier height increases monotonically with the coupling parameter  $\alpha$, and this effect is significantly amplified as $l$ increases, except for the case of $l=0$, as shown in Fig.~\ref{fig:veff}(c). However, when $\alpha=0.35$, the behavior of the potential peak as a function of $\beta$ changes from non-monotonic to monotonic and then back to non-monotonic as the angular momentum $l$ increases. The reason for the non-monotonic behavior at $l=0$ is the same as in QCM.

\subsection{Geodesic analysis}\label{sec:geo}

In this subsection, we study the geodesic behavior in the BH spacetime to compute the geometric absorption cross section and differential scattering cross section, and the glory approximation, which serve as comparisons with the partial-wave results.

For the geodesic motion of massive particles in the EsGB BH, we start from the classical Lagrangian
\begin{align}
    \mathcal{L}=\frac{1}{2}g_{\mu\nu}\dot x^\mu\dot x^\nu,\label{eq:cl_L}
\end{align}
where \(\dot x^\mu=dx^\mu/d\tau\) with \(\tau\)  being the proper time.
Using the metric \eqref{eq:metric} together with the timelike normalization \(g_{\mu\nu}\dot x^\mu\dot x^\nu=-1\), the following equations of motion can be derived from \cref{eq:cl_L}
\begin{subequations}
\label{eq:eoms}
\begin{align}
    \dot t&=\frac{E}{\mu \ee^A},\label{eq:t}\\
    \dot \varphi&=\frac{L}{\mu r^2},\label{eq:varphi}\\
    \dot r^2&=\frac{1}{\ee^B}\left(\frac{E^2}{\mu^2 \ee^{A}}-\frac{L^2}{\mu^2 r^2}-1\right),\label{eq:r}
\end{align}
\end{subequations}
where $L$ and $E$ are the conserved orbital angular momentum and energy of the particle, respectively.

\subsubsection{Geodesic absorption cross section}\label{sec:geo_s}

By introducing  the impact parameter $b = L/(vE)$, where $v = \sqrt{1 - \mu^{2}/E^{2}}$ denotes the asymptotic velocity of the particle, the radial equation \eqref{eq:r} can be rewritten as
\begin{align}
    \dot r^2=\frac{1}{(1-v^2)\ee^Br^2}\left\{\frac{r^2}{\ee^A}[1-(1-v^2)\ee^A]-v^2b^2\right\}.
\end{align}
When $\dot{r}^2=0$, there is a turning point $r_0$, where the radial velocity vanishes, and the test particle reverses its radial motion. If this turning point becomes degenerate, a critical radius $r_c$ is obtained, corresponding to a critical impact parameter $b_c$. For $b<b_c$, the test particle is captured by the BH, while for $b>b_c$, it is scattered back to infinity. The critical case $b=b_c$ corresponds to an unstable circular orbit at $r=r_c$, separating capture from scattering trajectories. The critical radius $r_c$ corresponds to the photon sphere, $r_{\text{ps}}\equiv r_c \approx 1.5\,r_h$ for a Schwarzschild BH. From Eq. \eqref{eq:eoms}, the conditions determining the critical radius and impact parameters in our EsGB metric are
\begin{align}
&\frac{r_c^2}{\ee^{A(r_c)}}[1-(1-v^2)\ee^{A(r_c)}]-v^2b_c^2=0,\\
&r_cA'(r_c)\ee^{A(r_c)}-2\ee^{A(r_c)}[1-(1-v^2)\ee^{A(r_c)}]=0.
\end{align}
For each BH solution obtained in~\cref{sec:background}, we can numerically determine the values of $b_c$ and $r_c$ from these conditions. 
Due to EsGB corrections, the critical radius is slightly shifted from that of the Schwarzschild BH. The critical impact parameter $b_c$ determines the absorption cross section of the BH as \cite{Decanini:2011xw}
\begin{align}
    \sigma_{\text{geo}}=\pi b_c^2.\label{eq:geo_abs}
\end{align}

\subsubsection{Classical differential cross section }\label{sec:cdcs}

Defining the variable $u=1/r$ and combining~\cref{eq:r,eq:varphi}, we can obtain the relation between $u$ and $\varphi$ as
\begin{align}    \left(\frac{\dd u}{\dd\varphi}\right)^2=\frac{1}{\ee^B}\left[\frac{1}{v^2b^2\ee^A}-u^2-\frac{1-v^2}{v^2b^2}\right]\equiv h(u).
\end{align}
A test particle with an impact parameter $b$ that is scattered back to infinity has a deflection angle
\begin{align}
    \Delta \theta(b)=2\int_0^{u_0}\frac{du}{\sqrt{h(u)}}-\pi,
\end{align}
where $u_0=1/r_0$. The relationship between the deflection  angle $ \Delta \theta$ and the scattering angle $\theta(b)$  reads
\begin{align}
    \theta(b,N)=|\Delta\theta-2N\pi|\in[0,\,\pi],\quad N=0,1,2,\ldots,
\end{align}
where $N$ denotes the number of times the particle orbits the BH before being scattered to infinity. The classical differential scattering cross section is then ~\cite{P.A.Collins_1973}
\begin{align}
    \frac{\dd\sigma_\text{cls}}{\dd\Omega}=\sum_N\frac{b(\theta,N)}{\sin\theta}\left|\frac{\dd b}{\dd\theta}\right|,\label{eq:geo}
\end{align}
where $b(\theta,N)$ is the inverse function of $\theta(b,N)$. 
\begin{figure*}[htp!]
 	\centering
\includegraphics[width=0.47\textwidth]{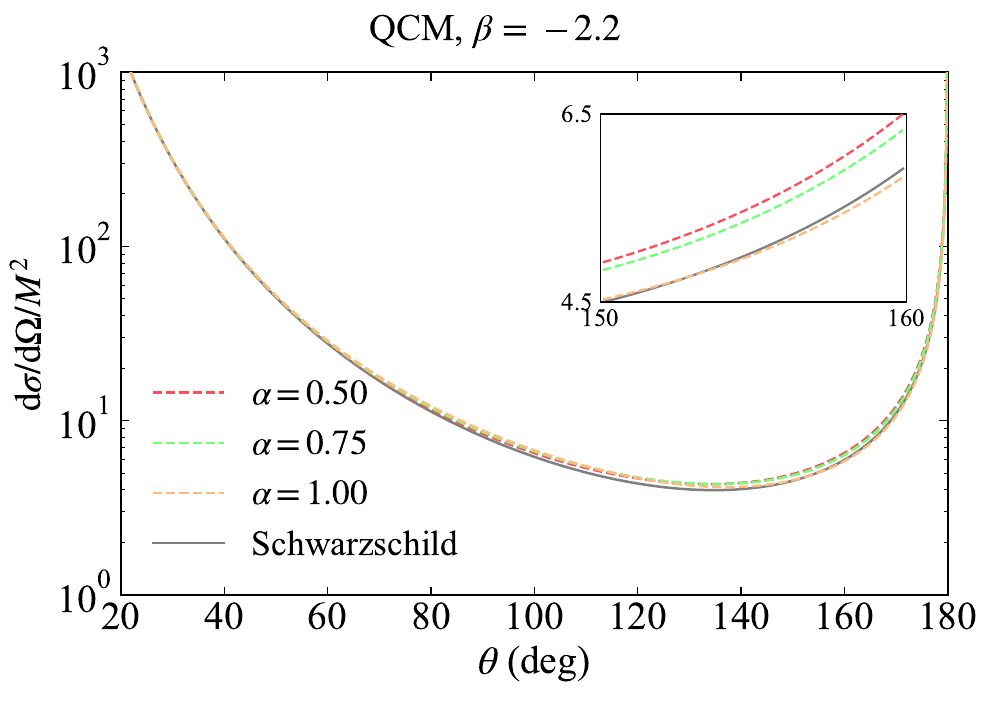}
\hfill
\includegraphics[width=0.47\textwidth]{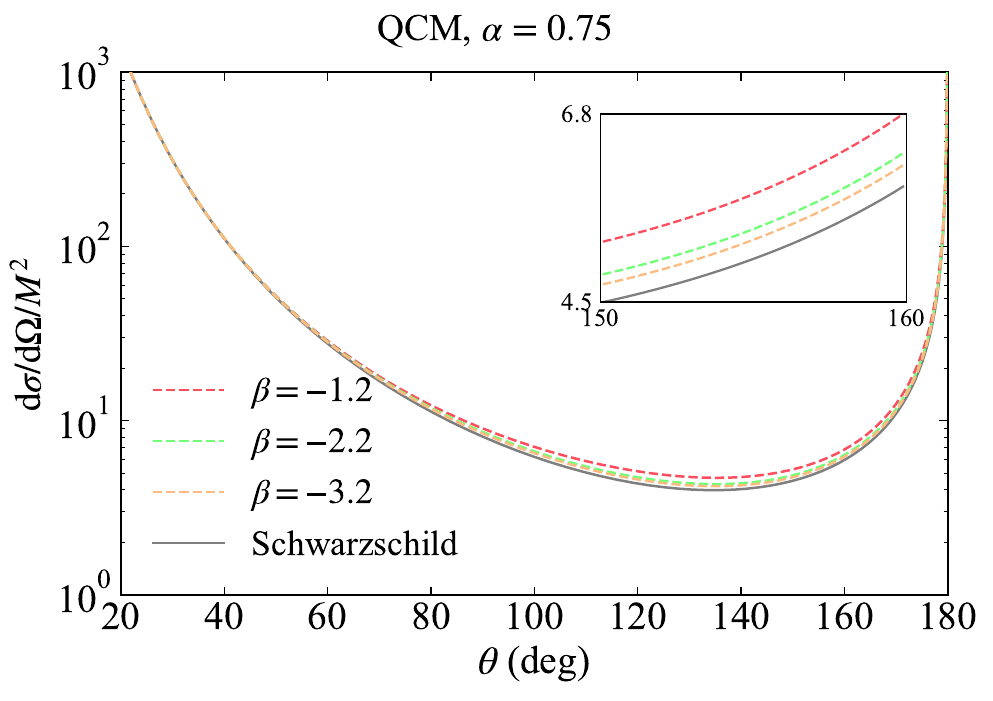}
\\
\includegraphics[width=0.47\textwidth]{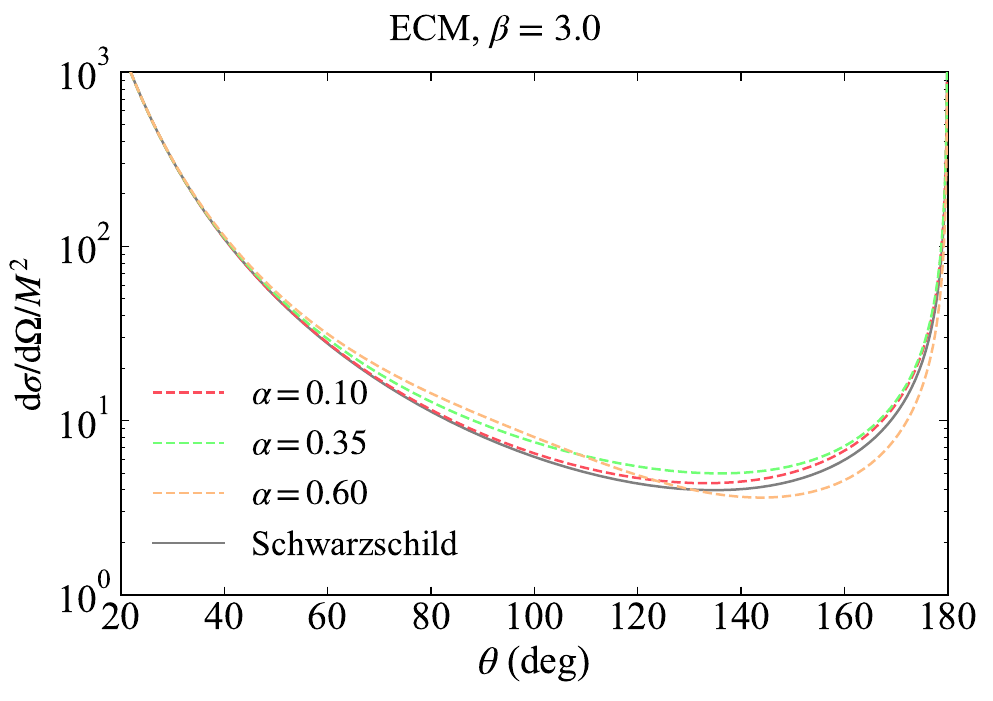}
\hfill
\includegraphics[width=0.47\textwidth]{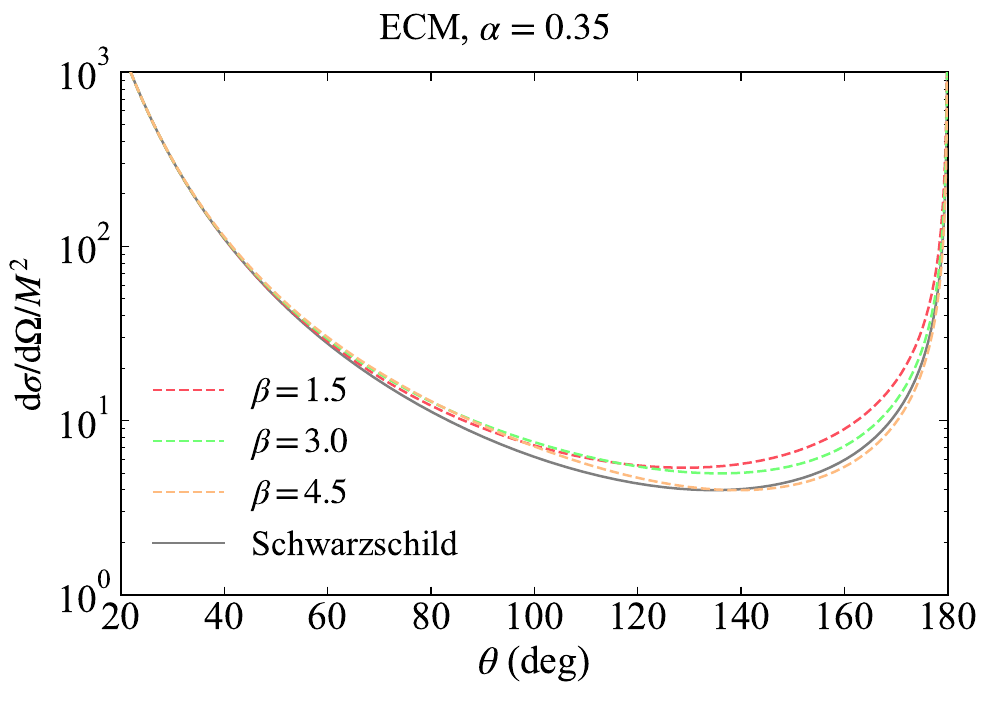}
  \caption{The classical differential cross sections for the QCM (top) and ECM (bottom) with different coupling parameters. Top left (a): $\alpha=0.5,\,0.75,\,1.00$ and $\beta=-2.2$; Top right (b): $\alpha=0.50$ and $\beta=-1.2,\,-2.2,\,-3.2$. Bottom left (c): $\alpha=0.1,\,0.35,\,0.6$ and $\beta=3.0$; Bottom right (d): $\alpha=0.35$ and $\beta=1.5,\,3.0,\,4.5$. The other parameters are fixed to $M\mu=0.4$ and $M\omega=3$.}
  \label{fig:geodiff}
\end{figure*}

We show the influence of the EsGB parameters on the classical differential scattering cross section in~\cref{fig:geodiff}. We observe that the EsGB parameters have little effect on the classical differential scattering cross section at small scattering angles. However, as the scattering angle increases, the classical differential scattering cross section gradually deviates from the Schwarzschild case, especially for ECM. This indicates that the EsGB parameters mainly modify the near-horizon spacetime structure, consistent with the larger metric deviation observed in that model.

\subsubsection{Glory scattering}

The classical absorption and differential scattering cross sections are based on the particle description of the scattering. In the $\theta\approx \pi$ limit, the wave nature of the incident matter causes the glory scattering effect. The glory differential cross section in this direction is given by \cite{PhysRevLett.52.2313,PhysRevD.31.1869} 
\begin{align}
\frac{\dd\sigma_\text{gl}}{\dd\Omega}=\sum_N 2\pi \omega v b(\pi,N)^2\left|\frac{\dd b}{\dd\theta}\right|_{\theta=\pi}\,J_0^2( \omega v b(\pi,N)\sin \theta),\label{eq:gs}
\end{align}

where $b(\pi,N)$ is the $N$-th impact parameter for the backscattered wave and $J_0$ denotes the Bessel function of the first kind of order zero. In~\cref{fig:glory}, we will use this result to check the validity of the partial-wave numerical results.

\section{Partial wave method absorption and scattering cross sections}
\label{sec:results}

In this section, we present and discuss the absorption and differential scattering cross sections obtained using the partial wave method from Eqs.~\cref{eq:abs,eq:dscs}.
To compute the cross sections, we numerically integrate~\cref{eq:radial_1} from near the horizon to large distances using the DOP853 method~\cite{Hairer1993}, and extract the reflection coefficients $R_{\omega l}$ by fitting the far-field numerical solution to the asymptotic form written in terms of spherical Hankel functions~\cite{PhysRevD.79.064014} through a linear least-squares procedure.
Due to the forward divergence at small angles, $\theta \approx 0$, we employ the series reduction technique described in Refs.~\cite{PhysRevD.74.064005,PhysRev.95.500} to resolve this issue.

\subsection{Absorption cross section}

\begin{figure*}[htp!]
\centering
\includegraphics[width=0.47\textwidth]{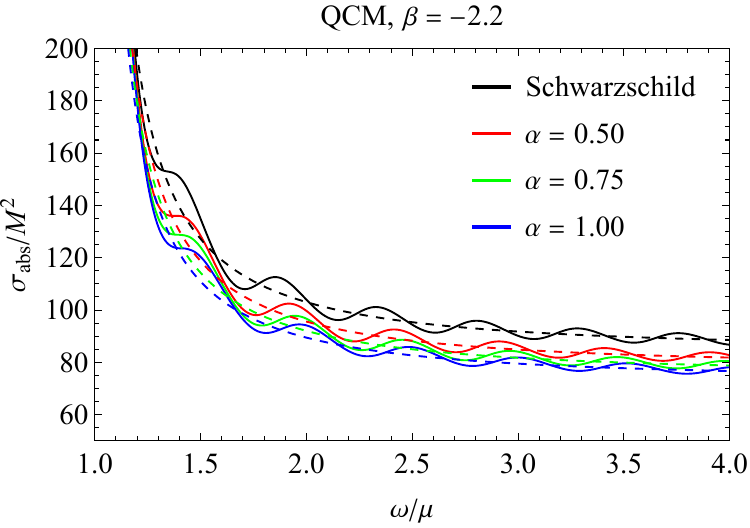}
\hfill
\includegraphics[width=0.47\textwidth]{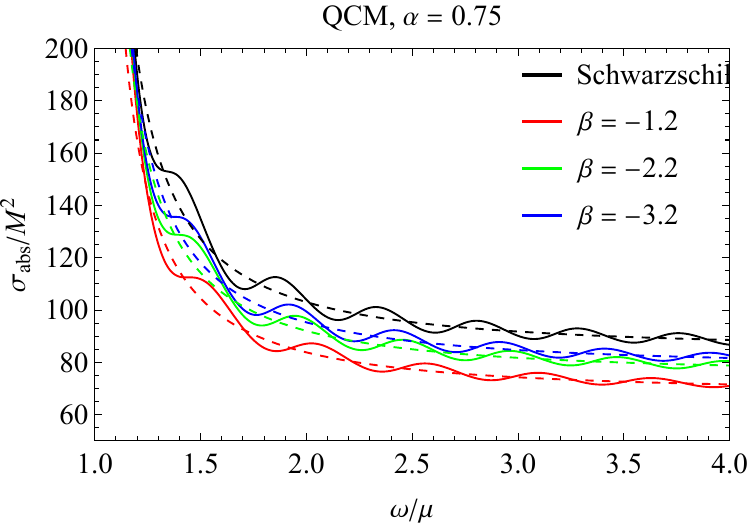}
\\
\includegraphics[width=0.47\textwidth]{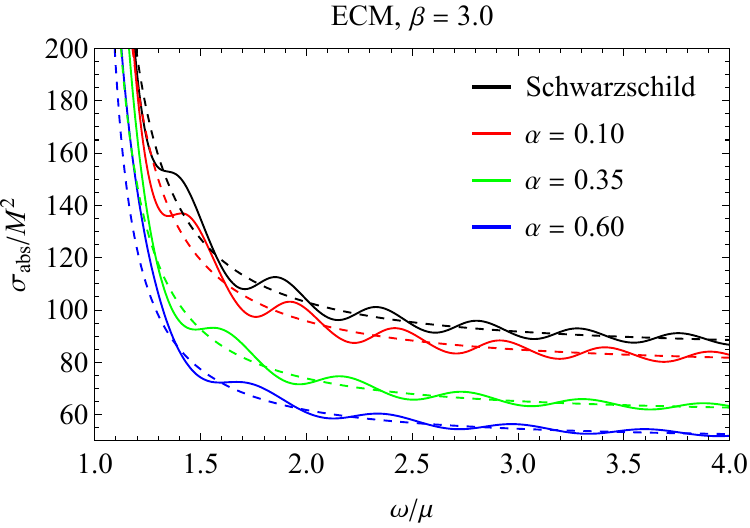}
\hfill
\includegraphics[width=0.47\textwidth]{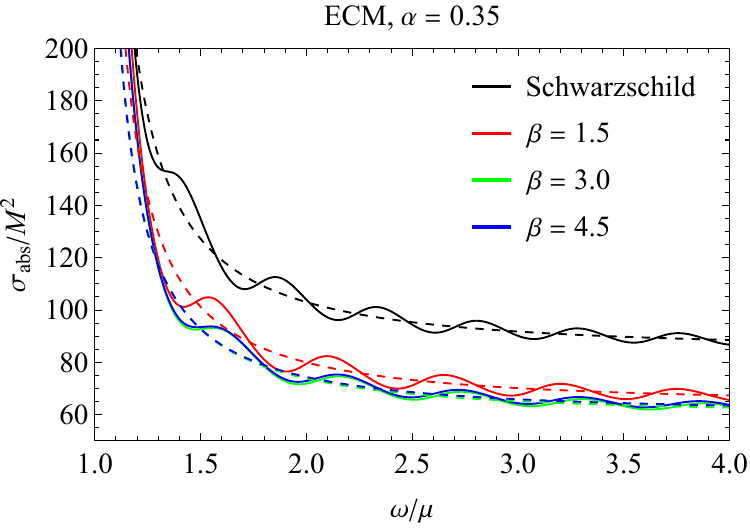}
\caption{The total absorption cross section for QCM (top) and ECM (bottom) with different coupling parameters. Top left (a): $\alpha=0.5,\,0.75,\,1.00$ and $\beta=-2.2$; Top right (b): $\alpha=0.75$ and $\beta=-1.2,\,-2.2,\,-3.2$. Bottom left (c): $\alpha=0.1,\,0.35,\,0.6$ and $\beta=3.0$; Bottom right (d): $\alpha=0.35$ and $\beta=1.5,\,3.0,\,4.5$. The other parameter is fixed to $\mu M=0.4$.}
\label{fig:abs}
\end{figure*}

\begin{figure*}[htp!]
	\centering
\includegraphics[width=0.47\textwidth]{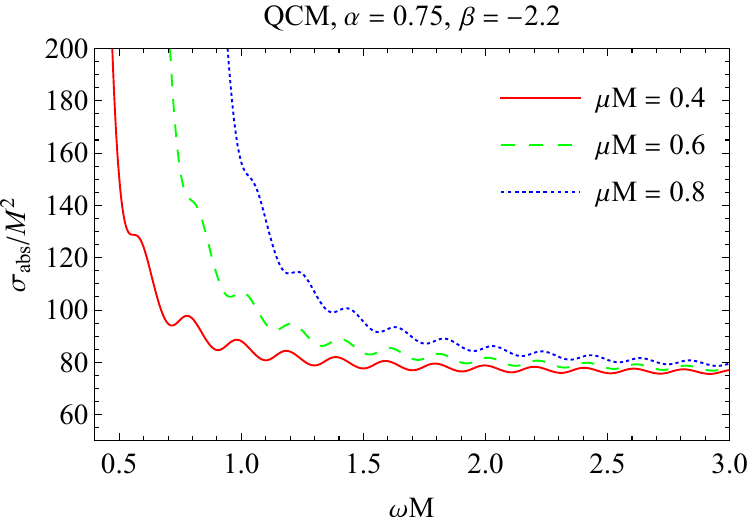}
\hfill
\includegraphics[width=0.47\textwidth]{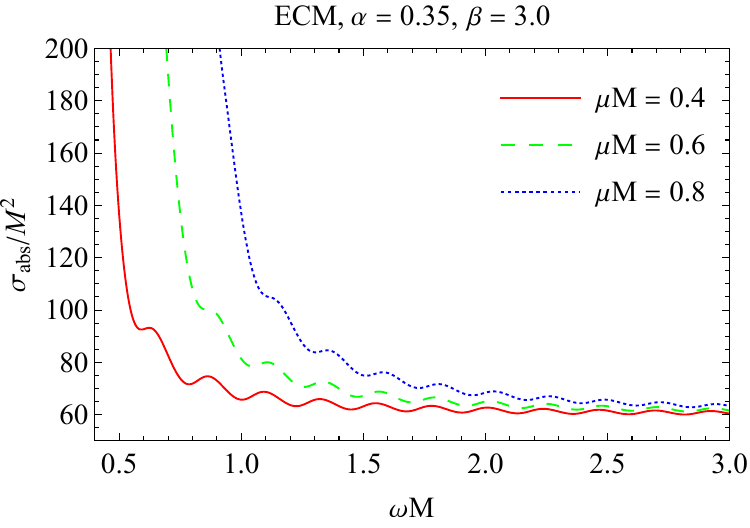}
	\caption{The total absorption cross section with different field mass \(M\mu\)  for QCM (left) with $\alpha=0.75,\beta=-2.2$ and ECM (right) with $\alpha=0.35,\beta=3.0$.}
	\label{fig:ab_muM}
\end{figure*}

In \cref{fig:abs}, we present the total absorption cross sections as a function of the wave frequency for the QCM and ECM with different coupling parameters $\alpha$ and $\beta$, and compare them with the geometric absorption cross sections obtained in~\cref{eq:geo_abs}. We observe that the total absorption cross section obtained from the partial wave method  (solid lines) oscillates around the geometric cross section (dashed lines) and gradually approaches the latter in the high-frequency limit. In addition, we observe that as $\omega/\mu\to 1$, the absorption cross sections of both Schwarzschild and EsGB BHs diverge. This is a universal feature of  the massive scalar field, arising from the gradual decrease of the asymptotic velocity $v=\sqrt{1-\mu^2/\omega^2}$. 

As for the specific effects of the parameters, we find from \cref{fig:abs} (a) and (c) that for both the QCM and ECM the presence of the coupling parameter $\alpha$ suppresses the absorption of scalar fields by the BH. This can be understood from the fact shown in Fig. \ref{fig:veff} that $\alpha$ increases the height of the potential, thereby reducing the transmission of the incident waves. As the absolute value of the coupling parameter $\beta$ increases, it is seen from \cref{fig:abs} (b) that the absorption cross section for the QCM gradually increases and  approaches that of the Schwarzschild BH, which is consistent with the corresponding decrease of the potential barrier. For the ECM, we find from \cref{fig:abs}(d) that the absorption cross section increases in the low-frequency region $1<\omega/\mu\lesssim1.3$ as $\beta$ increases. This is because the low-frequency absorption is mainly dominated by the low $l$ partial waves, such as $l=0$ and $l=1$. However, as $l$ increases, the peak of the potential barrier no longer varies monotonically with $\beta$, but instead first increases significantly and then decreases slightly. Consequently, in the intermediate- and high-frequency regions, the absorption cross section exhibits a non-monotonic behavior as the coupling parameter $\beta$ increases, first decreasing and then increasing, while the overall variation gradually becomes weaker.

Fig.~\ref{fig:ab_muM} presents the dependence of the absorption cross section on the field mass for QCM and ECM. In both models, increasing the field mass leads to an enhancement of the total absorption cross section. This indicates that the qualitative influence of the field mass on the absorption process is independent of the choice of the coupling function.  This behavior is consistent with our previous results~\cite{w9c5-53x3}. For scalar waves with the same incident frequency, increasing the field mass raises the effective potential barrier, which tends to suppress the absorption process. At the same time, however, the factor $1/k_\infty^2$ appearing in the absorption cross section also increases. The enhancement associated with this factor outweighs the suppressing effect of the higher potential, leading to an overall increase in the total absorption cross section.

\subsection{Differential scattering cross section}
\begin{figure*}[htp!]
	\centering
\includegraphics[width=0.47\textwidth]{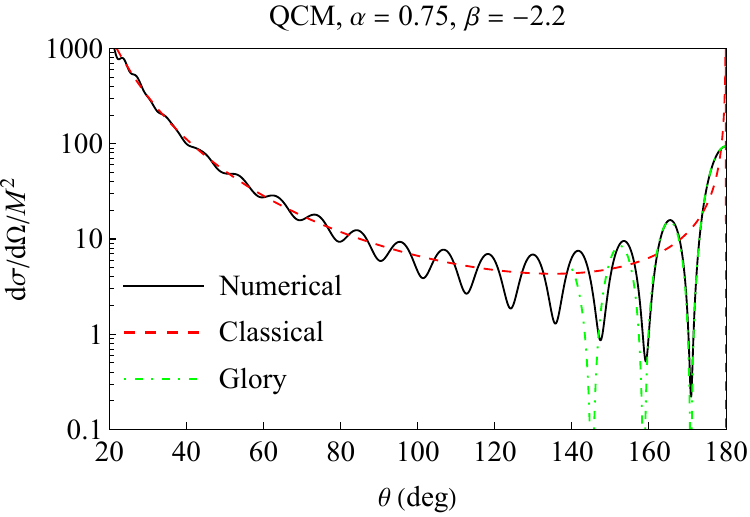}
\hfill
\includegraphics[width=0.47\textwidth]{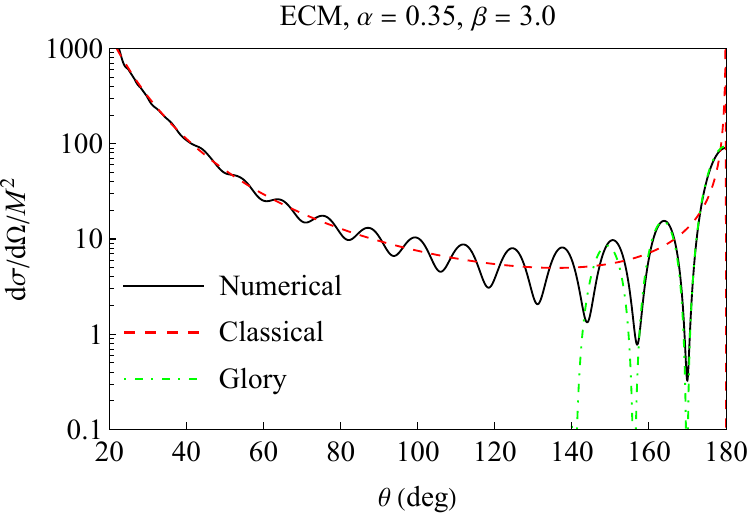}
	\caption{Comparison between the partial wave and the classical differential cross sections as well as the glory scattering for the QCM (left) with $\alpha=0.75,\,\beta=-2.2$ and ECM (right) with $\alpha=0.35,\,\beta=3.0$. $M\omega=3$ and $ M\mu=0.4$ are used in both plots. }
	\label{fig:glory}
\end{figure*}

In \cref{fig:glory}, we compare the partial wave and classical differential scattering cross sections together with glory scattering for a selection of parameters for the QCM ($\alpha=0.75$, $\beta=-2.2$, $M\omega=3$) and ECM ($\alpha=0.35$, $\beta=3.0$, $M\omega=3$) to verify their consistency. We find that the scattering features of the numerical BH are consistent with our expectations, exhibiting typical BHs scattering features such as forward divergence, interference oscillations, and backward glory scattering. In addition, as can be seen, in the small-angle region, the partial wave results for both models are nearly identical to the classical results in the geometric limit. In the large-angle region, the glory approximation also agrees very well with the partial wave numerical results. These agreements verify the correctness of our numerical method in solving the partial wave radial equation with the metric known only numerically. 

\begin{figure*}[htp!]
	\centering
\includegraphics[width=0.47\textwidth]{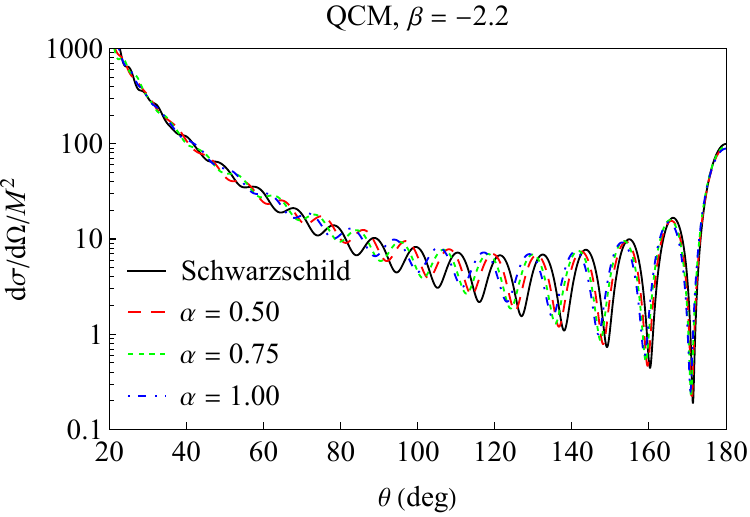}
\hfill
\includegraphics[width=0.47\textwidth]{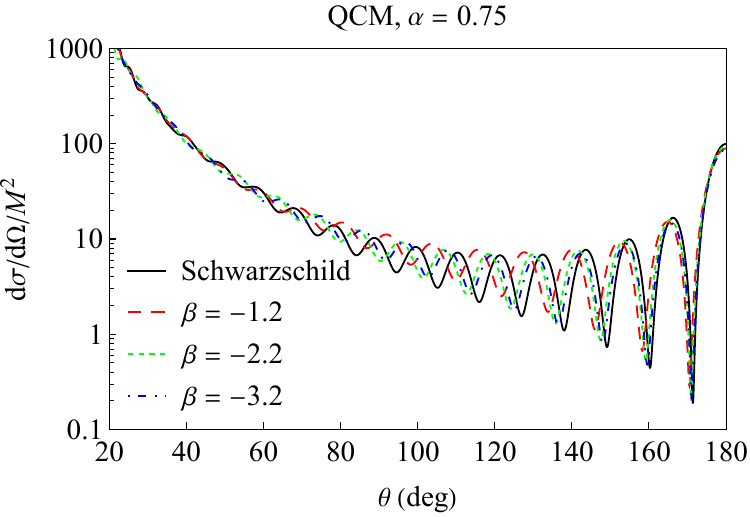}
\\
\includegraphics[width=0.47\textwidth]{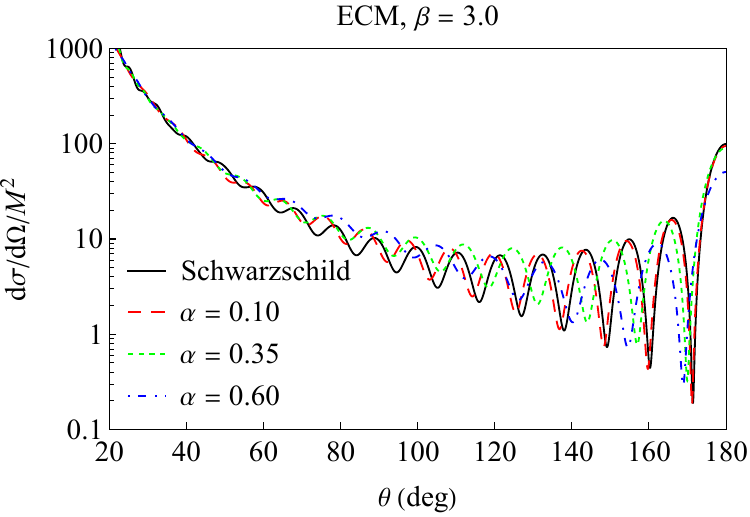}
\hfill
\includegraphics[width=0.47\textwidth]{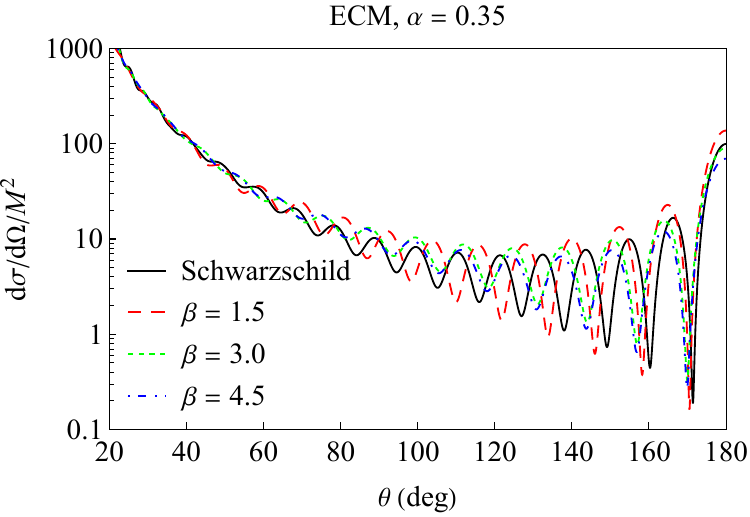}
	\caption{The differential scattering cross section for the QCM and ECM with different coupling parameters. Top left(a): $\alpha=0.5,\,0.75,\,1.00$ and $\beta=-2.2$; Top right(b): $\alpha=0.75$ and $\beta=-1.2,\,-2.2,\,-3.2$. Bottom left(c): $\alpha=0.1,\,0.35,\,0.6$ and $\beta=3.0$; Bottom right(d): $\alpha=0.35$ and $\beta=1.5,\,3.0,\,4.5$. The other parameter is  fixed to $M\mu=0.4$ and $M\omega=3$.}
	\label{fig:dsigma}
\end{figure*}

We present the effect of parameters $\alpha$ and $\beta$ for both the QCM and ECM on the differential scattering cross sections in \cref{fig:dsigma}. For the QCM, we clearly observe from the top row that the width of the interference fringes increases as the parameter $\alpha$ increases and $|\beta|$ decreases. This behavior can be understood in the following way. First we note from Eq. \eqref{eq:gs} it is the argument of the $J_0$ function, i.e. $\omega v b(\pi,N)$, that determines how the differential cross sections oscillate, including the fringe width, in the backward direction. Since $M\omega$ and $v$ are constants, the fringe width near the glory peak then is inversely proportional to the impact parameter $b(\pi,N)/M$, which is also extremely close to the critical $b_c/M$. On the other hand, from the geometric absorption cross section shown in Fig. \ref{fig:abs}, we knew that $b_c/M$ decreases as $\alpha$ increases and $|\beta|$ decreases in the QCM. This behavior indicates that $b(\pi,N)/M$ also decreases in this case, resulting in broader interference fringes.
However, for the effect of the parameter $\alpha$ in the ECM, \cref{fig:dsigma} (c) shows that this parameter not only modifies the width of the interference fringes but also significantly changes the oscillation amplitude. Specifically, the width of the interference fringes increases and the oscillation amplitude decreases as $\alpha$ increases. This behavior is similar to the effect of a stronger repulsive electrostatic interaction when a charged wave is scattered by charged BHs~\cite{5l97-88j1}.
For the effect of $\beta$, we observe from Fig. \ref{fig:dsigma} (d) that the oscillation amplitudes increase first and then decrease as $\beta$ increases, while the fringe width continues increase, although the increase is much slower when $\beta$ is large. 

\begin{figure*}[htp!]
\centering
\includegraphics[width=0.47\textwidth]{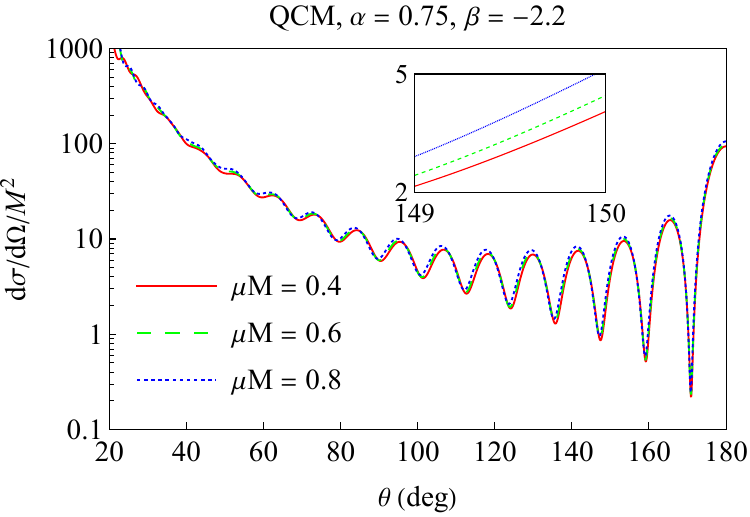}
\hfill
\includegraphics[width=0.47\textwidth]{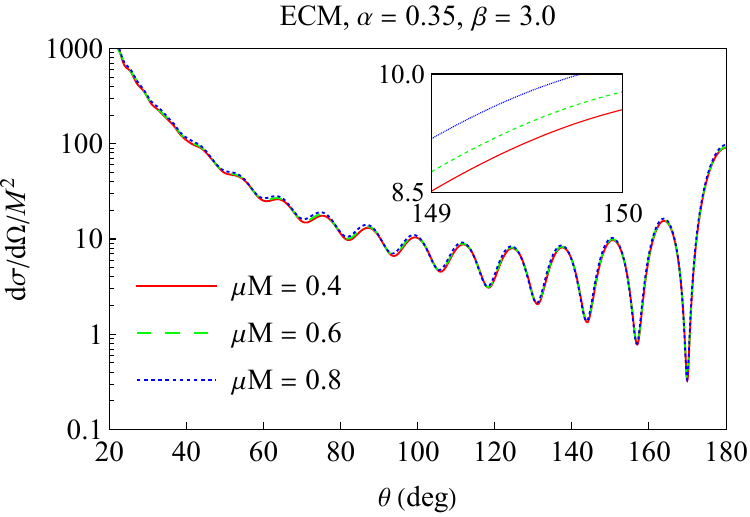}
  \caption{The differential scattering cross section with different field masses $M\mu$ for QCM (left) with $\alpha=0.75,\beta=-2.2$ and ECM (right) with $\alpha=0.35,\beta=3.0$. The other parameter are fixed at $M\omega=3$.}
  \label{fig:dsigma_muM}
\end{figure*}

In addition, we also studied the effect of field mass $M\mu$ on the differential scattering cross sections in both models in \cref{fig:dsigma_muM}. As $M\mu$ increases, the width of the interference fringes becomes wider, although very weakly for the given parameter range, in agreement with our previous results \cite{w9c5-53x3}. Again, this behavior can be explained by the decrease in the $v b(\pi,N)/M$ term in the argument of $J_0$ with increasing $M\mu$. 

\section{Conclusion}\label{sec:conclusion}

We have investigated the scattering and absorption of massive scalar fields by spontaneously scalarized BHs in both QCM and ECM in the EsGB gravity. The metrics of these BHs, although stable, are only obtained numerically via a Chebyshev pseudospectral method. Fortunately, a barycentric interpolation with spectral accuracy for the metrics still allows our application of the partial wave analysis for the cross sections. These partial wave results are cross-validated against their geodesic counterparts and the glory approximations, with excellent agreement found.

In general, it is found that the scalarization systematically suppresses the total absorption cross section, which can be attributed to the inward shift of the effective potential peak and the associated reduction of the critical impact parameter \(b_c/M\). The coupling parameters \(\alpha\) and \(\beta\) affect the scattering cross section in distinct ways in both models. Increasing $\alpha$ widens the fringes in both models, while increasing $\beta$ also widens them, although the broadening becomes much slower at larger $\beta$ in the ECM. More importantly, the differential scattering cross section reveals a qualitative distinction between coupling models: deviations from the Schwarzschild case in the QCM are concentrated in the intermediate-angle region \(\sim(70^{\circ}, 140^{\circ})\) with the glory peak nearly unchanged, whereas the ECM exhibits deviations across all angles including a significantly modified glory peak—in some parameter regimes (e.g., \(\beta = 1.5\)) even exceeding the Schwarzschild value. This angular fingerprint offers a potential observational diagnostic for discriminating among EsGB coupling functions.

Natural extensions of this work can be made in a few directions. The first is to extend the scattering study to axisymmetric EsGB BHs~\cite{KostasGlampedakis_2001}. It is expected that superradiance will happen in this case. The second line is to consider the scattering of gravitational waves by such BHs. Although the computations would be more involved due to the extra polarizations of the wave, the results might be more directly applicable to relevant gravitational wave observations. Lastly, we might explore the scattering of other BHs that are only known numerically, provided these BHs are stable and of certain importance. We are exploring some of these directions. 

\FloatBarrier

\bibliography{references}

\end{document}